\documentclass{emulateapj}

\makeatletter

\usepackage{graphics,graphicx,epsfig,lscape}

\slugcomment{}
\shorttitle{He-rich thermonuclear bursts from 4U 1728-34 }
\shortauthors{Misanovic, Galloway,  \& Cooper}

\makeatother

\begin{document}

\title{Ignition column depths of helium-rich thermonuclear bursts from 4U 1728-34}

\author{ Zdenka\ Misanovic$^{1}$, Duncan\ K.\ Galloway$^{1}$, and Randall\ L.\ Cooper$^{2}$}

\affil{$^{1}$Centre for Stellar and Planetary Astrophysics,
Monash University, Melbourne, VIC 3800, Australia \\
$^{2}$Kavli Institute for Theoretical Physics, University of California, Santa Barbara,\\ CA 93106, USA}

\begin{abstract}

We analysed thermonuclear (type-I) X-ray bursts observed from the low-mass X-ray binary 4U~1728$-$34 by {\it RXTE}, {\it Chandra}\/ and {\it INTEGRAL}. We compared the variation in burst energy and recurrence times as a function of accretion rate with the predictions of a numerical ignition model including a treatment of the heating and cooling in the crust.
We found that the measured burst ignition column depths are
significantly below the theoretically predicted values,
regardless of the assumed thermal structure of the neutron star interior.
While it is possible that the accretion rate measured by {\it Chandra}\/ is underestimated, due to additional persistent spectral components outside the sensitivity band, the required correction factor is typically 3.6 and as high as 6, which is implausible. Furthermore, such underestimation is even more unlikely for {\it RXTE}\/ and {\it INTEGRAL}, which have much broader bandpasses.
Possible explanations for the observed discrepancy include
shear-triggered mixing of the accreted helium to larger column depths, resulting in earlier ignition, or the fractional covering  of the accreted fuel on the neutron star surface. 

\end{abstract}


\section{Introduction}
\label{intro}


Thermonuclear (type I) bursts  are
triggered by unstable nuclear burning of the material accreted onto the
neutron star (NS) surface in low-mass X-ray binary (LMXB)
 systems  
\citep[e.g.,][]{1995xrbi.nasa..175L,2006csxs.book..113S}.
 The basic theory of type I bursts was outlined shortly after their detection
\citep[e.g.,][]{1976Natur.263..101W,1977Natur.270..310J,1977xbco.conf..127M,1978ApJ...220..291L}. According to these models,
 the accreted material, which usually consists
 of hydrogen and helium,
accumulates as  a thin layer on the NS surface (typically on time-scales
of hours to days), and when the pressure and temperature at its base reach
 critical values, the fuel will ignite and burn unstably until
exhausted.

Further observations \citep[see][for a review]{2008ApJS..179..360G} and
 subsequent modeling \citep{1981ApJ...247..267F,1987ApJ...319..902F,1987ApJ...323L..55F,1998mfns.conf..419B,2003ApJ...599..419N,2004ApJS..151...75W,2006ApJ...652..584C}
 showed in detail how the
burst properties depend on the accretion rate,  composition of the
accreted material (the H/He fraction and CNO metallicity), and  internal
 properties of the  neutron star.
For example, for systems accreting mixed H/He, the heat
generated from hydrogen burning  is usually a dominant factor for ignition. On the other hand, in evolved  systems
in which little or no H is present but the fuel consist mainly of He,
  the heat required for the burst ignition must come entirely from the 
electron captures and pycnonuclear reactions in the NS  crust.
Comparisons of He-bursts with ignition models thus offer a powerful
probe of the physical conditions in the neutron star crust, below the fuel
layer, and the cooling processes in the core \citep{1987ApJ...319..902F,2006ApJ...646..429C}.

 The best-known He-accretor 
 is a low-mass binary system 4U~1820$-$30, in which the neutron star orbits
its companion once every 11.4 minutes \citep{1987ApJ...312L..17S}.
 Such a tiny orbit cannot accommodate
a H-rich companion, and the mass donor is likely to be a He-rich white
dwarf \citep{1986Natur.323..105K}. 4U~1820$-$30 is in a bursting mode
 for around 40
days after switching to the low state \citep{2001ApJ...563..934C}, while
during the rest of its $\approx$176-day accretion cycle
\citep{1984ApJ...284L..17P} the source does not exhibit bursts \citep[][and references therein]{1984ApJ...282..713S}.
In a 20-hour EXOSAT observation during the low state, 
\citet{1987ApJ...314..266H} observed nearly regular bursting from 4U~1820$-$30, detecting  seven bursts 
 with a mean recurrence time of $3.21\pm0.04$ hours and persistent
 luminosity  of $L_X=2.8\times 10^{37}$ ergs s$^{-1}$ between the bursts.

 The burst ignition conditions for this source were modelled by
 \citet{2003ApJ...595.1077C}, who compared the predicted burst properties with
 the measurements of \citet{1987ApJ...314..266H}.
\citet{2003ApJ...595.1077C} presented models for pure He fuel, but also
estimated the effect of adding a small amount of hydrogen -- 5$-$35\% by mass
-- as predicted by some stellar evolutionary models
 \citep[e.g.,][]{2002ApJ...565.1107P}.
 The amount of energy released in pycnonuclear and electron
 capture reactions that escapes from the surface ($Q_{\rm crust}$) is a free
 parameter in this model, which also assumes the time-averaged rather than
 instantaneous accretion rate to set the crust temperature profile, because
 the thermal time in the crust is much longer than
 the $\approx$176-day accretion
 cycle. \citet{2003ApJ...595.1077C} found a good agreement between the model,
 which assumes a mixed fuel (10\% of hydrogen), and the data, providing that
 $Q_{\rm crust}=0.1$
 MeV/nucleon  \citep{2000ApJ...531..988B} and the time-averaged accretion rate
 is $\approx$2 times larger than the measured rate.
However, for a fuel consisting entirely of He, the required $Q_{\rm crust}$
was 0.4 MeV/nucleon and the required time-averaged accretion rate was
 4--5 times larger.  Self-consistent results were obtained with improved burst ignition models, in which
  the flux flowing outwards was calculated directly  from the neutron star crust and core neutrino emissivity and the core thermal
 conductivity \citep{2006ApJ...646..429C}.   
\citet{2006ApJ...646..429C} concluded that, in order to produce the bursts
separated by $\approx$3 hours, the core neutrino emissivity must be very
inefficient (e.g., suppresed modified Urca process) and the accretion rate must be $\approx$2 times larger than that inferred from the X-ray luminosity.

Although all  previous studies of He-bursts have focused on 4U~1820$-$30,
 the intermittent occurrence of the
 bursts makes
triggering of burst observations  extremely difficult, and only a few recurrence times and corresponding accretion rate estimates are available. 
A much more suitable candidate for such studies is the source 4U~1728$-$34,
 which
consistently exhibits frequent bursts characteristic of pure He fuel.
There is a total of 106 bursts from this source in the {\sl RXTE} burst catalogue
\citep{2008ApJS..179..360G}. The $\alpha$ values ($\approx$200), short rise times and decay time scales suggest a He-rich fuel. The persistent flux during {\sl RXTE} observations was 
$1-7 \times 10^{-9}$ erg cm$^{-2}$ s$^{-1}$ while the burst recurrence times
were on average $\approx$4 hours. 
Evidence for a short orbital period of 10.77~min has been detected recently in
 the analysis
of {\sl Chandra} observations (Galloway et al.~2010; in prep.),
  supporting the long-suspected
 identification of 4U~1728$-$34  as an ultracompact LMXB. 
 The source is probably accreting pure He from its evolved companion
and is practically a twin of 4U~1820$-$30,
 except for the much more frequent and
 reliable bursting. 4U~1728$-$34 has also been observed extensively
 by {\sl INTEGRAL}
\citep{2006A&A...458...21F,2006AstL...32..456C}. 
\citet{2006A&A...458...21F} detected 36 type I bursts during the transition from hard to soft state, where the source luminosity increased from 2$-$12\% of the Eddington luminosity.

In this paper we present analysis of  new {\sl Chandra} observations of 
4U~1728$-$34. We measured the recurrence times and corresponding accretion
rates of the 25 bursts detected  during a 240-ks HETGS exposure. The
detailed spectral analysis and the detection of the orbital period and radius
expansion bursts are reported in the companion paper (Galloway et al.~2010), while we present the comparison of the observed burst properties  with a  new ignition model. 
In Section~2 we present the data and describe our analysis of the {\sl Chandra} observations. The measured burst properties, which include the accretion rates (estimated  from the persistent flux), burst fluences, $\alpha$ values and burst recurrence times are presented in Section~3. The new burst ignition model is described in
 Section~4, while Section~5 shows the comparison of the model and data. Possible
 explanations of the discrepancy between the data and our ignition model are
 discussed in Section~6. Finally, our conclusions are summarized in Section~7.


\section{Observations and analysis}
\label{observations-and-results}
\subsection{Chandra observations}
\label{chandra}

4U~1728$-$34 was observed between 2006 July 17$-$23 with the HETGS abroad
{\sl Chandra}. The observations 6568, 6567 and 7371 were made in the continuous
clocking (CC) mode with exposures of 49.5, 151.8 and 39.7 ks, respectively. A
total of 25 type I bursts, separated by 1.8$-$3.9 hours, were detected in
these observations (for the light curve and more details on the {\sl Chandra}
observations see Galloway et al.~2010). Galloway et al. (2010) also report on
the detection of a period of 10.77 min in the low-energy persistent intensity (which they interpret as arising from orbital modulation) and describe a search for line emission and photoionization edges  in the persistent spectra between the four radius expansion bursts detected in observation 6568.
In this paper we focus on measuring the pre-burst accretion rates and corresponding burst recurrence times, and their comparison with our new ignition models.

The first order HEG and MEG persistent spectra  were
extracted using  time intervals starting 150 seconds after the previous
burst, and ending 50 seconds before the burst peak, to exclude any possible
 burst
emission. These spectra, which contained approximately 2000 to 3000 total
counts, were binned to a minimum of 100 counts per bin,
 and fitted simultaneously for each interval.
The combination of the CC mode (selected to minimize photon pileup) and grating observations is non-standard, making the correct background subtraction possible only for the brightest sources \citep[e.g.,][]{2001AJ....122...21M}. Hence, we restricted our spectral  fitting  to the 1.5$-$6 keV band, in which the contribution from the background is minimal. The response matrices were produced by the CIAO task {\sc mkgrmf}, while the task {\sc fullgarfs} was used to produce the auxiliary files, which were used for fitting both the persistent and time-resolved burst spectra.

Since the {\sl Chandra} energy band is relatively narrow, to help us select
the appropriate spectral model, we examined spectral fitting results of previous observations made in significantly broader bands.  
 The source was in a soft state during the broadband (0.1$-$100 keV) {\sl
   BeppoSAX} observation by \citet{2000ApJ...542.1034D}. The best-fit model
 consisted of two components, a 2-keV blackbody (with an emitting region
 comparable to the expected radius of the neutron star), and a Comptonized component ({\sc comptt} model in {\sc xspec}) with seed photon temperature of $\approx$1.5 keV, electron temperature of $\approx$ 10 keV and optical depth of $\approx$5. \citet{2000ApJ...542.1034D}  also detected two broad emission lines at $\approx$6.7 keV and $\approx$1.6 keV, probably emitted in the ionized corona.
\citet{2006A&A...448..817D}  reported on the analysis of the simultaneous {\sl RXTE} and {\sl Chandra} observations of 4U~1728$-$34. The broadband (1.2$-$35 keV) spectrum was best described by  a blackbody ($kT \approx$0.6 keV) and a Comptonized component
 ($kT_{0}\approx$1.5 keV; $kT_{\rm e}\approx$7 keV; $\tau\approx$5), but no emission lines were detected. Instead, \citet{2006A&A...448..817D} have found absorption edges at $\approx$7 keV and 
$\approx$9 keV associated with Fe{\sc i} and Fe{\sc xxv}.      
 During several {\sl INTEGRAL} observations (in the 3$-$200 keV band), the source was found undergoing the transition from  the intermediate/hard to soft state with the electron temperatures of the Comptonizing plasma decreasing from $\approx$35 keV to 3 keV \citep{2006A&A...458...21F}.

Following these results,
we selected a single-component\footnote{
A  blackbody component may also be present when the source is in the soft state, in which case the estimated bolometric flux would be reduced (by at most $\approx20$\% for our data). By including a blackbody component the contribution from the comptt component is reduced, and the latter contributes more to the total bolometric flux, particularly at high energies. }
absorbed {\sc comptt} model in {\sc XSPEC} to
fit the HEG and MEG data.
We then used the spectral results from previous observations to fix some of the model parameters as the rather limited {\sl Chandra} band was not sufficient to constrain them. First, we adjusted and fixed the seed photon energy to 0.4 keV for the first four bursts detected in the observation 6568, 0.5 keV for the 18 bursts in  6567 and 0.3 keV in the last three bursts detected in the observation 7371.
These values were selected so that the corresponding absorption column was
consistent with the best-fit value found from the burst spectra ($N_H=2.29
\times 10^{22}$ cm$^{-2}$, see below), although we actually found that the
 measured fluxes were not very
sensitive to this parameter. Since the {\sl Chandra} band was well below the
spectral turnover ($\approx$10--20 keV, for $kT_{\rm e}\approx$3 keV), we could  select and fix a  wide range of electron temperatures and get statistically
 acceptable fits. 
We selected the temperature of 35 keV to
 represent a hard spectral state, and then repeated the fitting with a low (3 keV) plasma temperature typical of a soft state. Since the narrow {\sl Chandra} band was not sufficient to discriminate between these spectral states, we used both models to estimate the hard and soft-state persistent X-ray fluxes for each burst, from which we calculated the corresponding accretion rates.

\begin{figure}
\begin{center}
\includegraphics[height=8cm,angle=0]{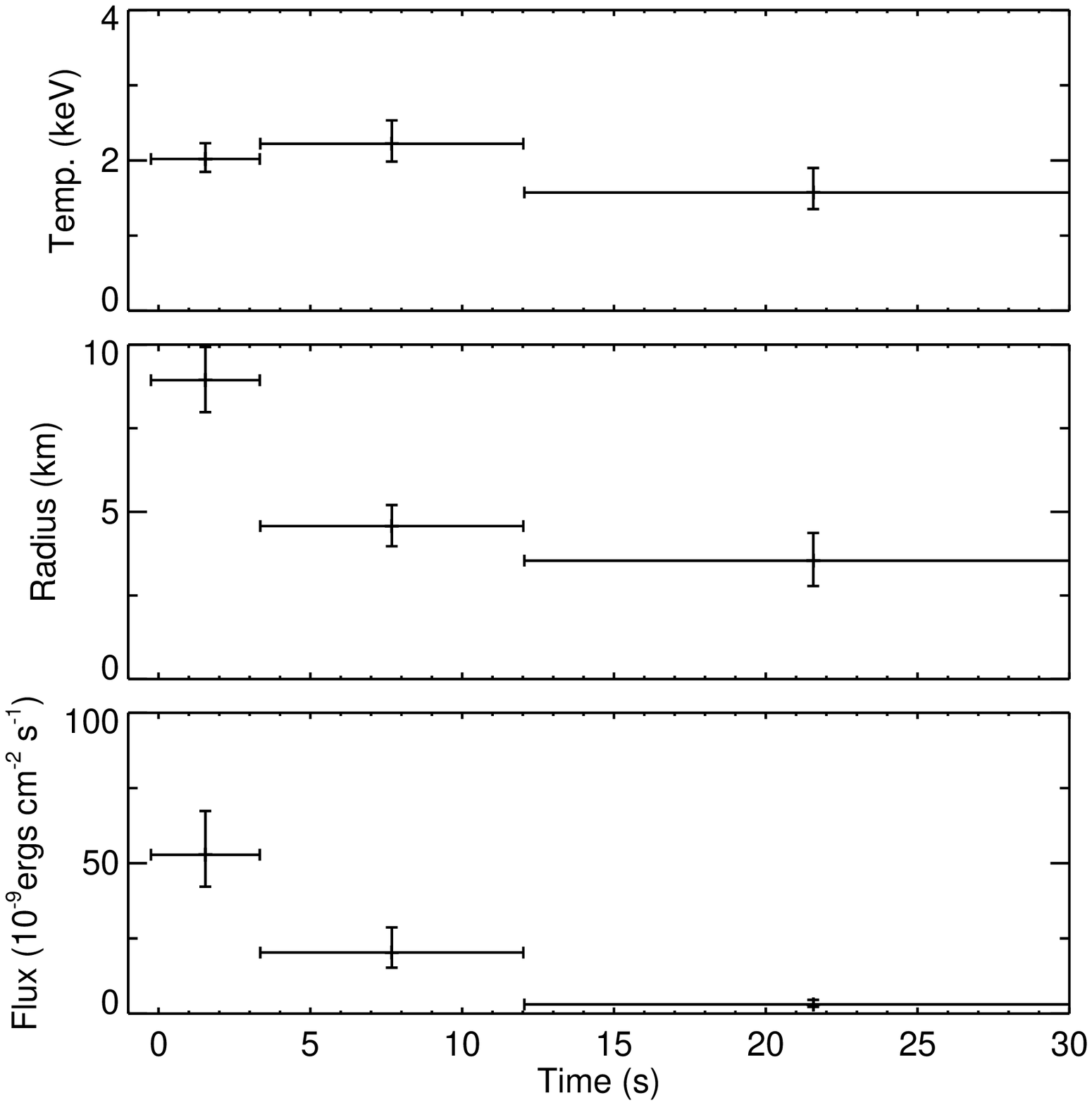}
\includegraphics[height=8cm,angle=0]{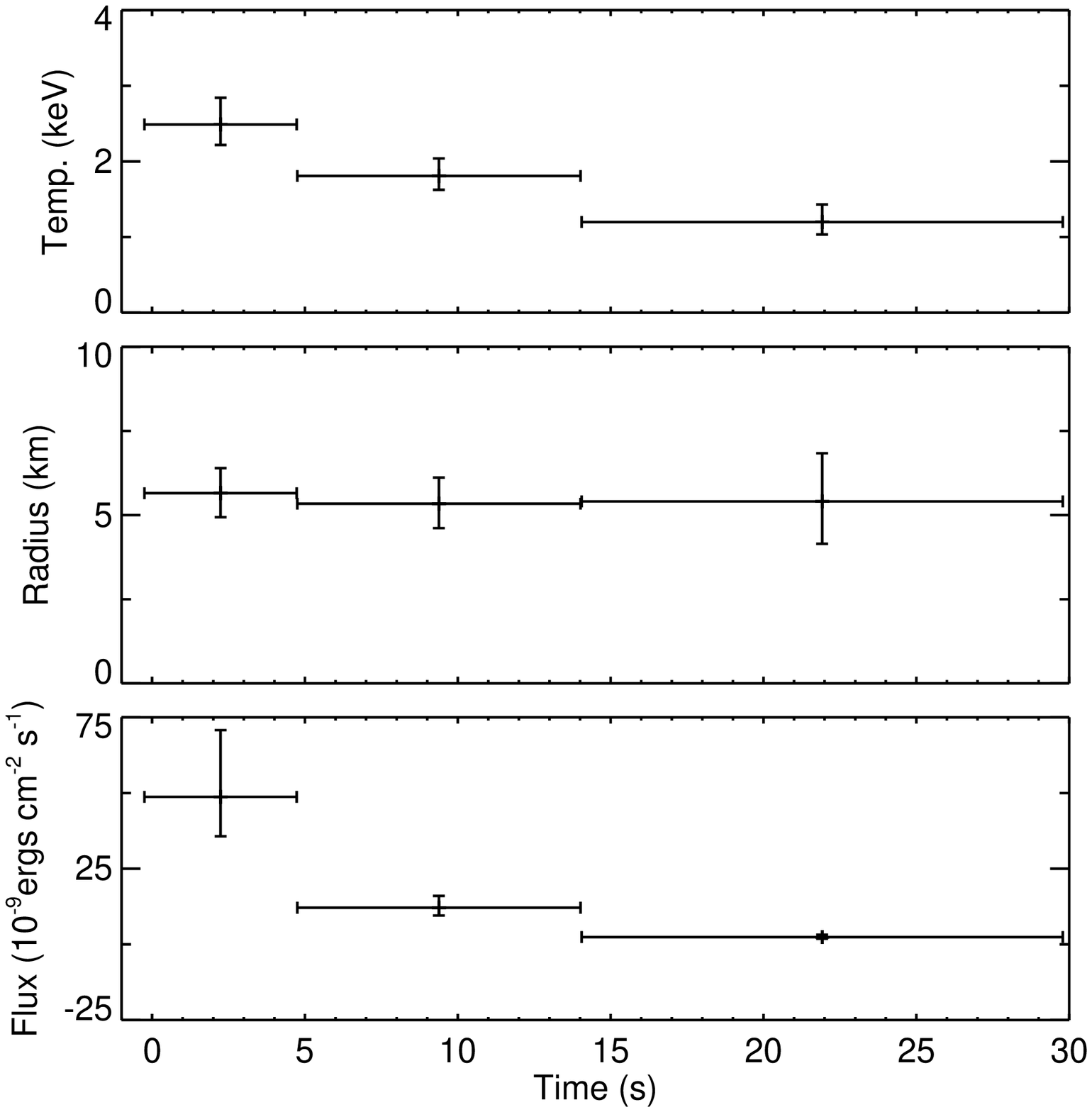}
\end{center}
\caption{The fitted blackbody temperature, radius (assuming a distance of 5.2 kpc) and bolometric flux
  measured during the bursts \#1 (top) and \#5 (bottom) in the {\sl Chandra} data. The error bars indicate the 1 $\sigma$ uncertainties.  Each time-bin contains $\approx$800 cts. The burst peak occurs 1$-$1.5 sec from the start of the burst (t=0 sec). 
\label{fig-time-resolved-spectra}}
\end{figure} 

\begin{figure}
\begin{center}
\includegraphics[height=8cm,angle=0]{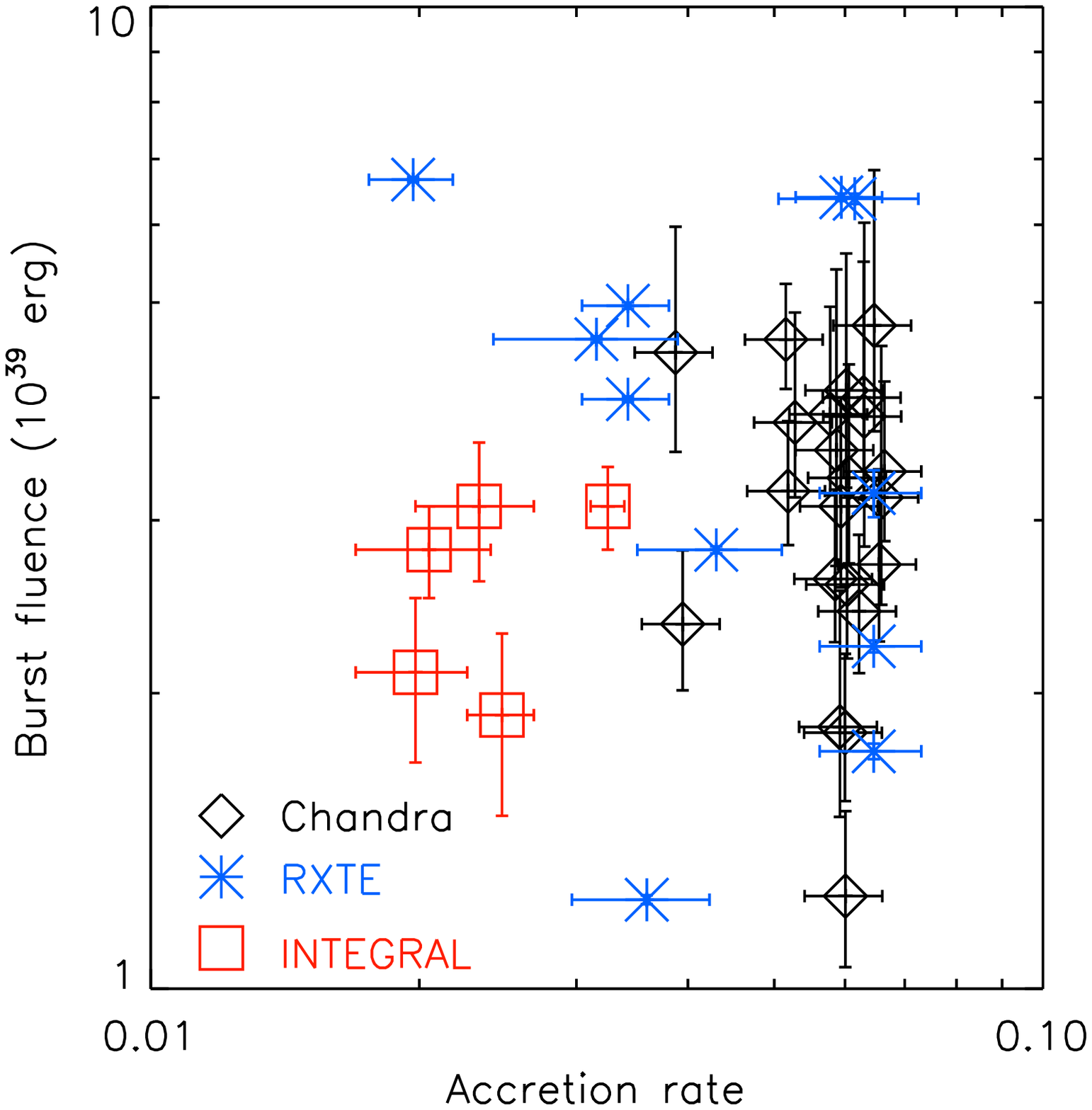}
\includegraphics[height=8cm,angle=0]{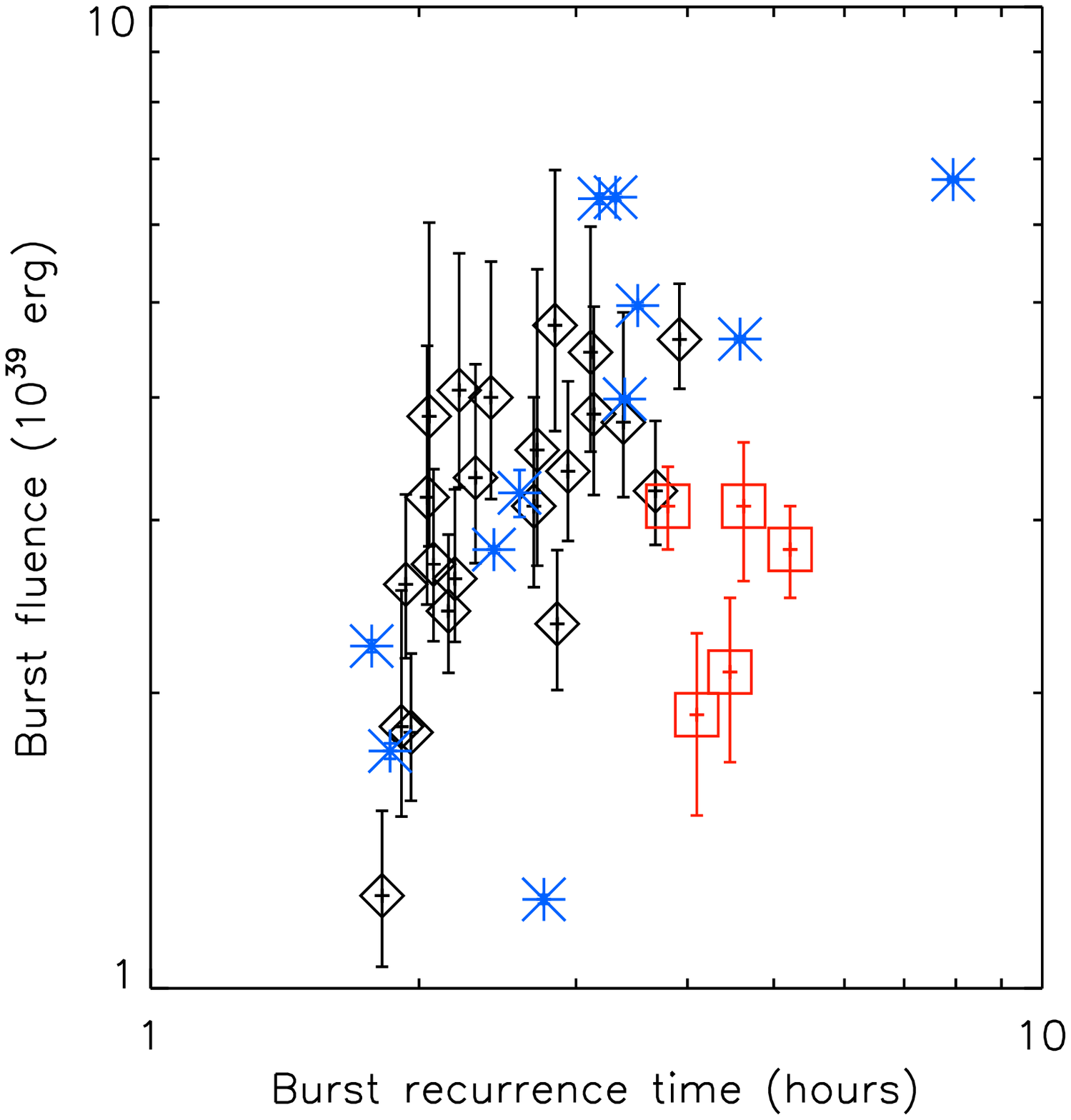}
\end{center}
\caption{{\bf Top:} The burst fluence versus the accretion rate
  measured from the persistent X-ray luminosity and
 expressed in units of
  the Eddington luminosity, assuming the accreted material to be pure
  helium. {\bf Bottom:} The burst fluence versus the recurrence time. 
\label{fig-fluences}}
\end{figure} 

To measure  the neutral hydrogen column density, we first
produced the peak spectra of all bursts using only the data around the peak
(between 0.5 s before the peak and 1.5 s after the peak)
 and combined these
spectra for each observation. Since Galloway et al. (2010) showed  that the
burst profiles in each observation have approximately the same shape
 (see their Fig.~3), we
combined the peak spectra of the four bursts from the observation 6568,
eighteen bursts from 6567 and three bursts from 7273. We subtracted the
background (persistent) emission from these combined spectra and fitted them
with  absorbed blackbody models. We obtained good fits for each of the
three observations with an $N_{H}=2.30^{+0.32}_{-0.29} \times 10^{22}$ cm$^{-2}$ in the first
observation,  $N_{H}=2.27^{+0.48}_{-0.44}\times 10^{22}$ cm$^{-2}$ in the
 second, and
$N_{H}=2.31^{+1.40}_{-1.13}\times 10^{22}$ cm$^{-2}$ during the last three
 bursts. We then
calculated the weighted mean and variance of $N_{H}=2.29^{+0.26}_{-0.24}\times 10^{22}$ cm$^{-2}$.

Finally, we produced time-resolved spectra for each individual burst by
applying an adaptive time-binning, i.e., by accumulating the counts until a
 minimum of
 approximately 800 counts (total HEG and MEG first order) in each bin.
 We then fitted the HEG and MEG spectra of each bin
 with an
absorbed blackbody model (with the $N_H$ fixed at the mean value of $2.29\times 10^{22}$ cm$^{-2}$) and used the
 model parameters to calculate the
corresponding bolometric flux. The fluence of each burst was then calculated
by summing the measured fluxes over the burst. As one example, in
Fig.~\ref{fig-time-resolved-spectra} we show the best-fit blackbody model
 parameters
and fluxes for time-resolved spectra of two bursts from our sample.
Since the absorption column measurements from persistent spectra in previous observations
 range from   
$1.6-2.8 \times 10^{22}$ cm$^{-2}$, we  estimate the maximum
possible systematic error of $20-25$\%. We also note that for the {\sl RXTE} bursts selected for our analysis  (Section~\ref{rxte-integral-data}) \citet{2008ApJS..179..360G} performed the spectral fitting with  the absorption column fixed at the mean value for each burst. For the 11 {\sl RXTE} bursts included here, this value is in the range $0.6-2.4 \times 10^{22}$ cm$^{-2}$.

\subsection{RXTE and INTEGRAL bursts}
\label{rxte-integral-data}

From the  total of 106 bursts observed  from this source by  {\sl RXTE} and
re-analysed by \citet{2008ApJS..179..360G}, we selected  11 bursts for which
the recurrence times could be reliably measured. We used the catalogued values
of the persistent X-ray luminosities \citep[measured in the 2.5-25 keV band and
multiplied by the bolometric correction factor of 1.38;][]{2008ApJS..179..360G}, burst fluences and
$\alpha$ parameters.

 4U~1728$-$34 was in the field of view of JEM-X and ISGRI cameras during several thousand short {\sl INTEGRAL} observations of the Galactic centre region made in 2003 and 2004.  
A total of 36 bursts detected in the combined {\sl INTEGRAL} observations were catalogued by 
\citet{2006A&A...458...21F},
 while a complete analysis of all available {\sl INTEGRAL} IBIS  observations by  \citet{2006AstL...32..456C} revealed 61 bursts from the source. 
The burst recurrence times, however, could only be accurately measured  if the source was in the JEM-X field of view continuously between two subsequent bursts (M. Falanga, private communication), which we could  verify for 5 bursts. These 5 bursts, which we include in our analysis, are listed in the {\sl INTEGRAL} burst catalogue by \citet{2006A&A...458...21F}.
\citet{2006A&A...458...21F} measured
  the X-ray luminosities (in the 3$-$200
keV band) and used them to estimate the accretion rates. The
burst fluences were measured from the time-resolved burst spectra with a
typical  bin
size of 1$-$2 seconds.


\section{Measured burst properties}
\subsection{The persistent luminosity and accretion rates}
\label{persistent-vs-fluence}

Using the best-fit models of the persistent pre-burst spectra, we measured the
absorbed persistent fluxes  in the 1.5$-$6 keV band to be in the range $1.16-2.30 \times 10^{-9}$ ergs cm$^{-2}$ s$^{-1}$, and estimated the corresponding intrinsic bolometric luminosities\footnote{assuming a distance of 5.2 kpc measured from the photospheric radius expansion bursts in {\sl RXTE} data by \citet{2008ApJS..179..360G}.} ($1.3-2.3 \times 10^{37}$ ergs s$^{-1}$ for the $kT_{\rm e}$=3 keV, and $4.1-6.4 \times 10^{37}$ ergs s$^{-1}$ for the $kT_{\rm e}$=35 keV), from which we calculated the pre-burst mass accretion rates. 
The range of the estimated bolometric luminosities is approximately 10\% smaller than that of the fluxes in the 1.5-6~keV band, due to the different choice of $kT_0$ for the three observations (0.3, 0.4 and 0.5~keV for observation 7371, 6568 and 6567, respectively; see Section~2.1). Simple phenomenological (power-law) fits indicate that the persistent spectra actually become slightly harder with increasing flux, suggesting that the range in bolometric flux should be larger rather than smaller. However, these effects are relatively minor compared to the uncertainty in the electron temperature $T_e$.

The accretion rates calculated from the persistent flux,
assuming Comptonization in the high temperature plasma ($kT_{\rm e}\approx$35 keV),
would imply  almost one order of magnitude larger $\alpha$ parameters (Section~\ref{fluences}) than usually measured from this source.
 In addition, since the relatively high persistent
 flux during the {\sl Chandra} observations suggests a soft spectral state, we conclude that the accretion rates estimated from the 3-keV X-ray
 luminosities are more realistic, and we will assume these values throughout.

The accretion rates are expressed in units of the Eddington limit. 
The Eddington accretion rate for pure helium fuel is assumed
to be $1.78 \times 10^{18}$ g s$^{-1}$ in the distant observer's frame, which corresponds to the Eddington
luminosity  of $3.5 \times 10^{38}$ ergs s$^{-1}$
\citep[e.g.,][]{2003ApJ...599..419N}.
Although the ultra-compact systems are thought to accrete almost pure helium
(i.e., the fraction of the accreted hydrogen $X_{\rm 0}=0$), some stellar evolution models predict that the hydrogen fraction can be as high as 20\% \citep{2002ApJ...565.1107P}. To account for this possibility, we compared all  our calculations with the case $X_{\rm 0}=0.2$, where we used
the corresponding Eddington accretion rate of $1.48 \times 10^{18}$ g
s$^{-1}$ \citep[$L_{\rm Edd}=2.9 \times 10^{38}$ ergs s$^{-1}$;][]{2003ApJ...599..419N}.

The persistent accretion rate during the first two {\sl Chandra}
 observations was
nearly constant at around 0.066 of the Eddington limit, decreasing to
$\approx$0.038 during the last pointing, when the last three bursts occurred.
 The persistent accretion rate
 during the five
bursts detected with {\sl INTEGRAL} and eleven {\sl RXTE} bursts varied
between   approximately 0.02 
and 0.08 of the Eddington accretion rate ($\dot{M}_{\rm Edd.}$).

\subsection{Burst fluences and $\alpha$ parameters}
\label{fluences}

Fig.~\ref{fig-fluences} shows the burst fluences, measured from the
time-resolved burst spectra (Section~\ref{observations-and-results}),
 and plotted
against the burst accretion rate (left) and recurrence time (right).  
 The bursts are weak ($1-8\times 10^{39}$ erg) and we see a sharp increase of the burst energy with burst recurrence time, but only for short recurrence times ($\approx 2$ hours).

The parameter $\alpha$ is usually used to compare
the gravitational energy ($Q_{\rm grav}$) produced by accretion and the energy
released by nuclear burning ($Q_{\rm nuc}$) during the
burst \citep[$\alpha=zc^2/Q_{\rm nuc}$; where $z$ is the surface gravitational redshift
and $c$ is the speed of light; e.g.,][]{1987ApJ...319..902F}. The nuclear energy depends on the fuel composition and is 
calculated as $Q_{\rm nuc}=1.6+4\langle X \rangle$ MeV per nucleon,
 where the hydrogen
fraction in the fuel, $\langle X \rangle$, is averaged over the burning layer. The formula gives an energy
of 4.4 MeV per nucleon for the Solar abundance of hydrogen ($\langle X \rangle=0.7$), which
implies\footnote{The
  approximate formula, $\alpha \approx 44\,{\rm MeV/nuc}\,(M/1.4\,M_{\rm
        sun})(R/10\,{\rm km})^{-1}(Q_{\rm
        nuc}/4.4\,{\rm MeV/nucl})^{-1}$, is given by \citet{2008ApJS..179..360G}, who omitted
  the redshift correction factor $1+z \approx 1.3$.} $\alpha \approx 60$, while higher values suggest a He-rich fuel. 
We have calculated the $\alpha$ parameters for the {\sl Chandra} bursts as
$\alpha = F_{\rm p} \times \Delta t/E_{\rm b}$, where
$F_{\rm p}$ is the persistent flux integrated over the burst recurrence time $\Delta t$, and $E_{\rm b}$ is the burst
fluence, and found  it to be in the range  105$-$342 (the mean value 189),
which implies a He-rich fuel. Consistent $\alpha$ values are catalogued for the {\sl RXTE} (91$-$314) and {\sl INTEGRAL} bursts (160$-$240) with the mean value of 180 for the whole sample.

In addition to the $\alpha$ parameter, from the measured burst fluences ($E_{\rm b}$) we calculated the total mass burnt during the  burst
 ($M_{\rm b}=E_{\rm b}/Q_{\rm nuc}$; where $M_{\rm b}$ is measured in the distant observer's frame).
This allows us to compare the consumed mass during the burst directly to the mass accreted since the preceding burst.     
According to the thin-shell instability models, the accreted mass is burnt completely during the bursts, unless there are some energy ``leaks'', for example, due to stable hydrogen burning  before the burst or other sources of energy loss \citep[e.g.,][and references therein]{1987ApJ...319..902F}.

We calculated    $M_{\rm b}$ for the pure helium fuel, and repeated the calculation for  $X_{\rm 0}=0.2$. 
For the Solar metallicity,
this fraction of hydrogen will be completely burnt in $\approx$3 hours, via the
hot CNO cycle
\citep[e.g.,][]{2004NuPhS.132..435C}. Since most of the measured burst recurrence times are between 3 and 5 hours (Fig.~\ref{fig-fluences}), no hydrogen is left to burn unstably and  the
$Q_{\rm nuc}$ does not increase significantly for these bursts.
 However, if the metallicity is, for example, ten times lower
 ($Z=0.0012$), the
time needed to burn the accreted hydrogen is almost 32 hours, which means that
most of the accreted hydrogen will be left to burn unstably during the burst.
 As a consequence, $Q_{\rm nuc}$ will increase up to $\approx$\,2.3 MeV per nucleon for bursts with short recurrence times.

\begin{figure}
\begin{center}
\includegraphics[height=8cm,angle=0]{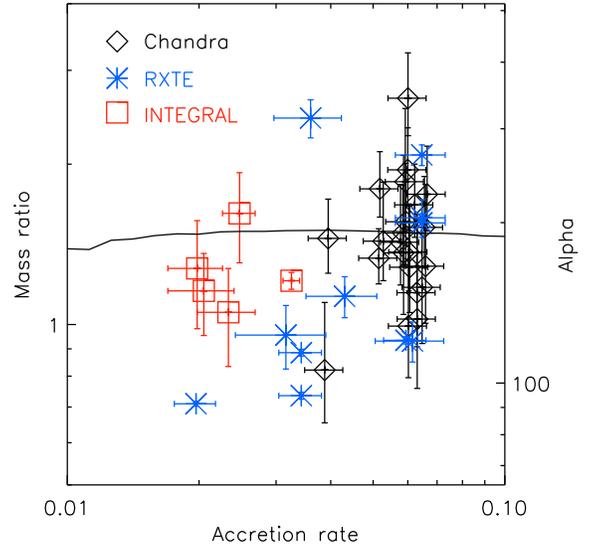}
\end{center}
\caption{The ratio of the accreted mass and the mass consumed during the burst  plotted as a function of the persistent accretion rate (in units of the Eddington limit, assuming the accreted material to be pure helium)
 for {\sl Chandra} (diamonds), {\sl RXTE} (asterisks) and {\sl INTEGRAL} data (squares). The corresponding $\alpha$ parameter is plotted on the right Y-axis.  The line shows the $\alpha$ values predicted by our model (see Section~5).
\label{fig-ratio-accRate}}
\end{figure} 

Fig.~\ref{fig-ratio-accRate} shows the ratios of the accreted and burnt mass and corresponding $\alpha$ parameters  versus the measured
accretion rate, assuming the fuel consists of pure helium.
The accreted mass is on average 1.4 times larger 
than the total mass of fuel burnt during the burst.
 However, there is a large scatter around this value, with the fraction of the
 exhausted fuel of  $\approx$50\% or less during several bursts.    
We also notice several  bursts with the total burnt mass that
appears to exceed the accreted mass (i.e., the mass ratio $<1$).  
 Mass ratios greater than
$\approx$2 are measured for three bursts, which also have large $\alpha$
 parameters
($\approx$300).
A small $\alpha$ value of $\approx$90 was measured for the two bursts with
the corresponding mass ratio of $\approx$0.7.


\section{Ignition models}
\label{ignition-models}

We use the general-relativistic global linear stability analysis of
\citet{2005ApJ...629..422C} to theoretically determine the type I X-ray burst
recurrence time and $\alpha$ value as a function of accretion rate.  We assume steady spherical accretion onto a neutron star of gravitational mass $M = 1.4 \; M_{\odot}$ and areal radius $R = 10.4 \; \mathrm{km}$ at a rate $\dot{M}$, where $\dot{M}$ is the rest mass accreted per unit time as measured by an observer at infinity.  We describe the accreted matter's composition by the hydrogen mass fraction $X$, metal mass fraction $Z$, and helium mass fraction $Y=1-X-Z$.

We make the following two modifications to the model of \citet{2005ApJ...629..422C}. (1) \citet{2005ApJ...629..422C} assumed the energy generated by electron captures, neutron emissions, and pycnonuclear reactions in the crust was distributed uniformly.  We now follow \citet{HaenselZdunik2008} and distribute the energy according to their Table A.3.  (2) We updated the crust thermal conductivity according to \citet{ShterninYakovlev2006} and \citet{ChugunovHaensel2007}.

The accretion rate, accreted matter composition, and interior temperature profile together determine the recurrence time.  The temperature profile is a strong function of the crust's thermal conductivity and core's neutrino emissivity, both of which are poorly constrained.   The thermal conductivity is a decreasing function of the impurity parameter $Q_{\mathrm{imp}} = \langle Z^{2} \rangle - \langle Z \rangle^{2}$ \citep{ItohKohyama1993,2009ApJ...703..994D}.   Although \citet{SBCW1999} found $Q_{\mathrm{imp}} \sim 100$ from steady-state nucleosynthesis calculations, subsequent investigations suggest $Q_{\mathrm{imp}}$ should be much smaller \citep{SBC2003,2004ApJS..151...75W,KHKF2004,HBB2007,HCB2009,SYHP2007,2009ApJ...698.1020B}.  
Also, the core neutrino emissivity, and thereby the core cooling rate, depends on the unknown ultradense matter equation of state \citep[for reviews, see][]{YakovlevPethick2004,PGW2006}.  
Unless noted otherwise, we set $Q_{\mathrm{imp}} = 100$ throughout the crust and adopt a slow,
suppressed core cooling model in which nucleon-nucleon bremsstrahlung
processes dominate the neutrino emission \citep[see, e.g., Table 1 of
][]{PGW2006}; these parameters are realistic limits on the crust thermal
conductivity and core neutrino emissivity and are set to  maximize the burst
ignition region temperature and thereby minimize the burst recurrence time, for the closest correspondence with observations.


\section{Comparison with observations}

\clearpage

\begin{figure}
\begin{center}
\includegraphics[height=6.5cm,angle=0]{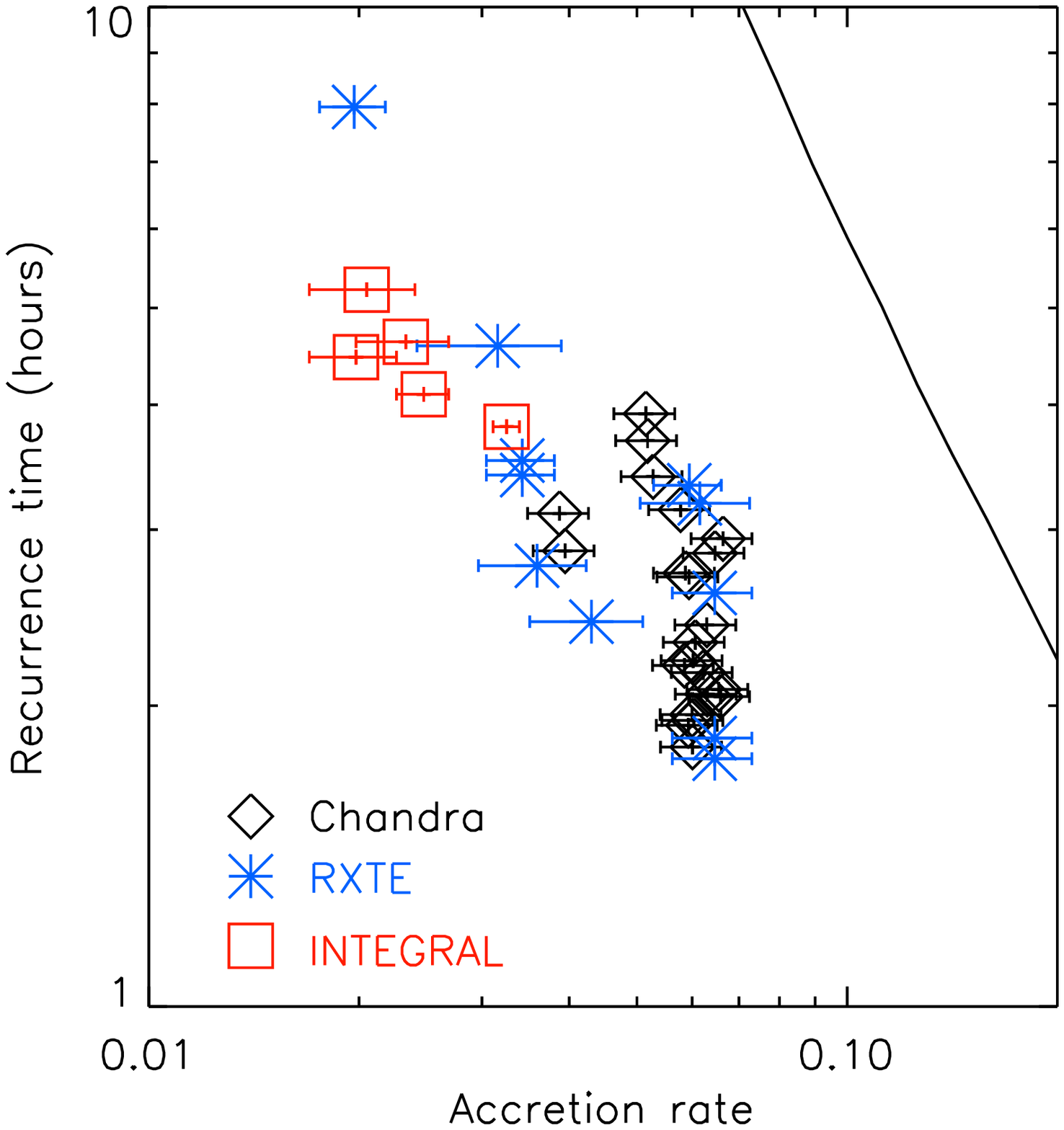}\includegraphics[height=6.5cm,angle=0]{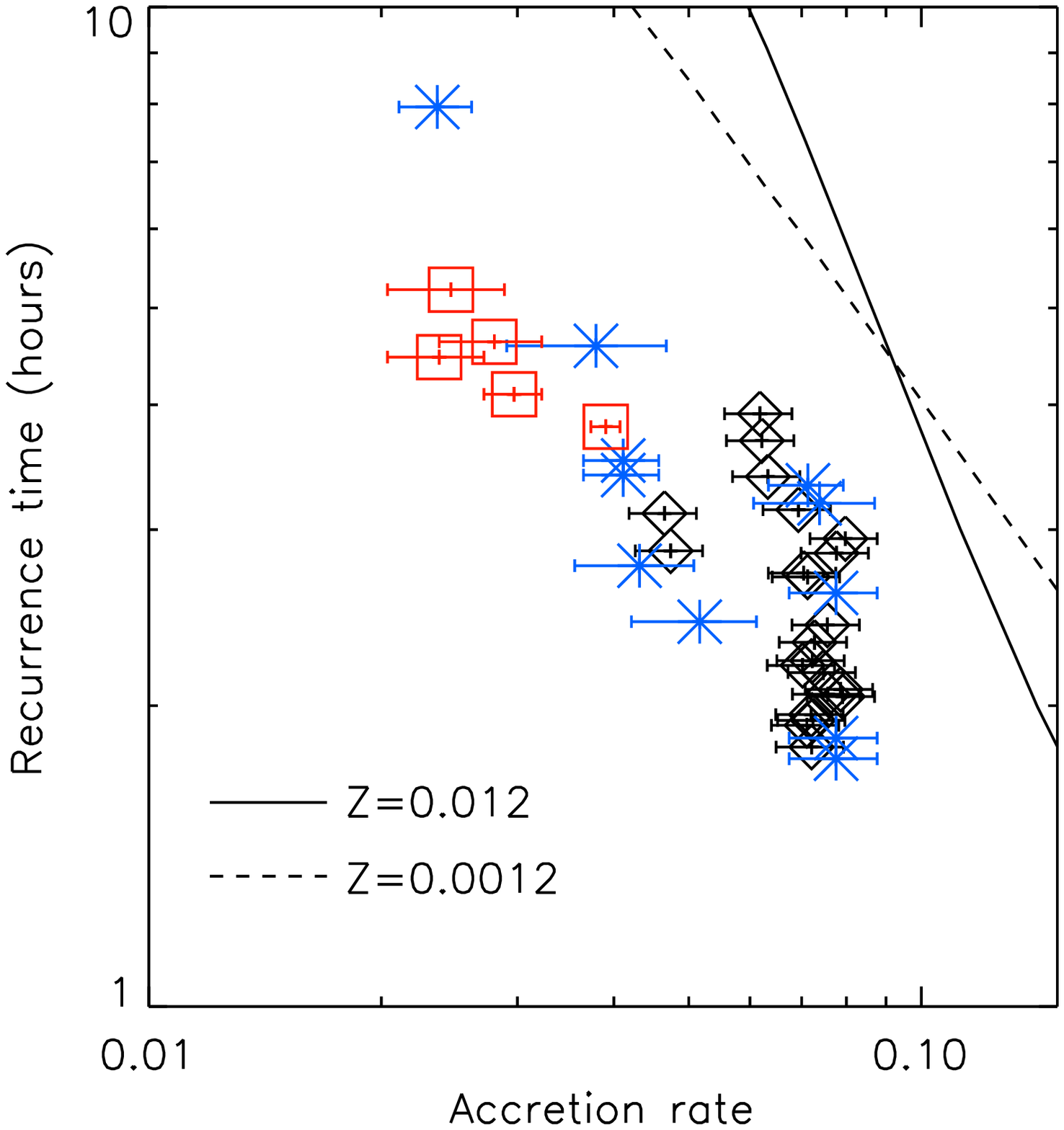}
\includegraphics[height=6.5cm,angle=0]{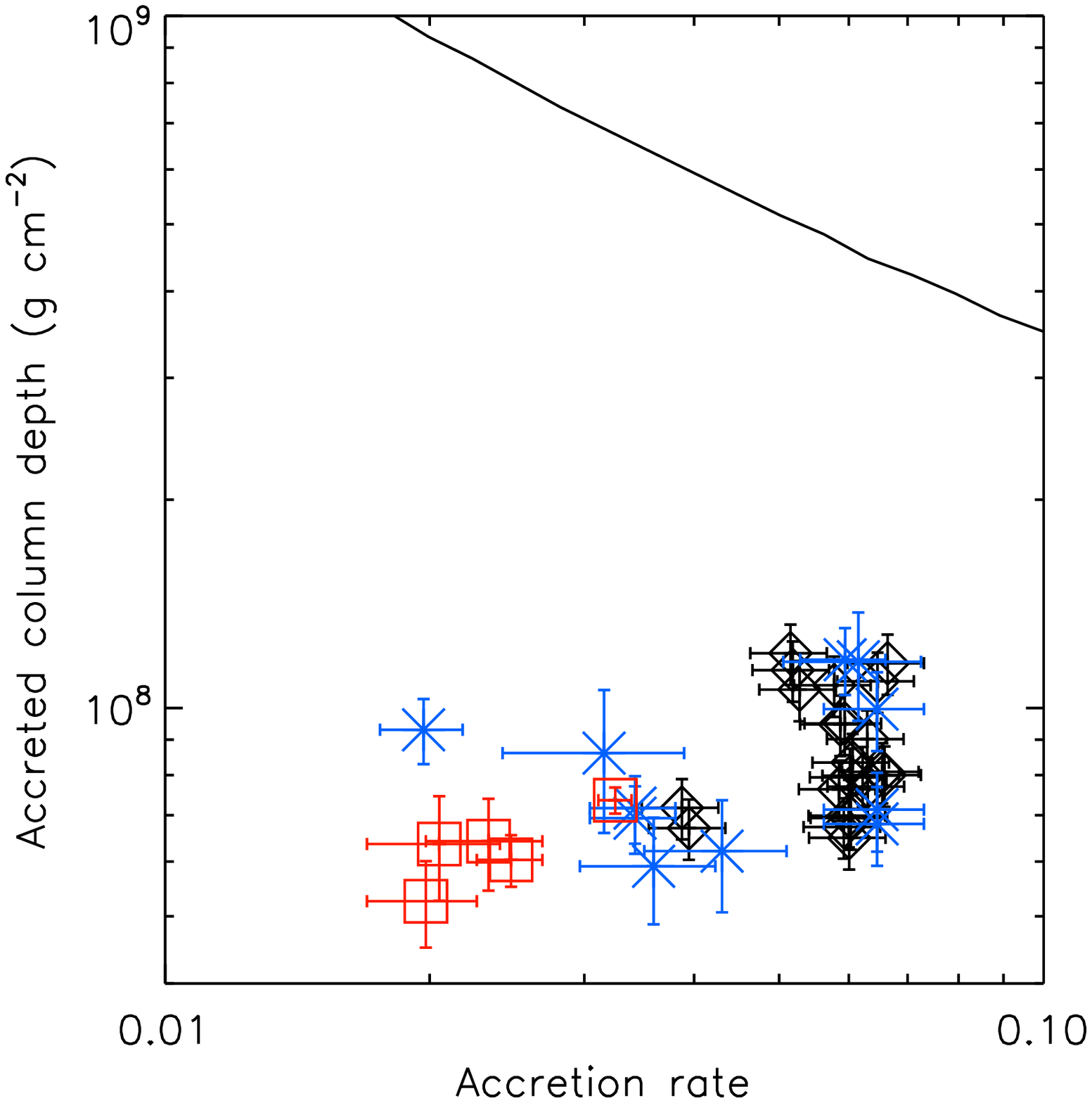}\includegraphics[height=6.5cm,angle=0]{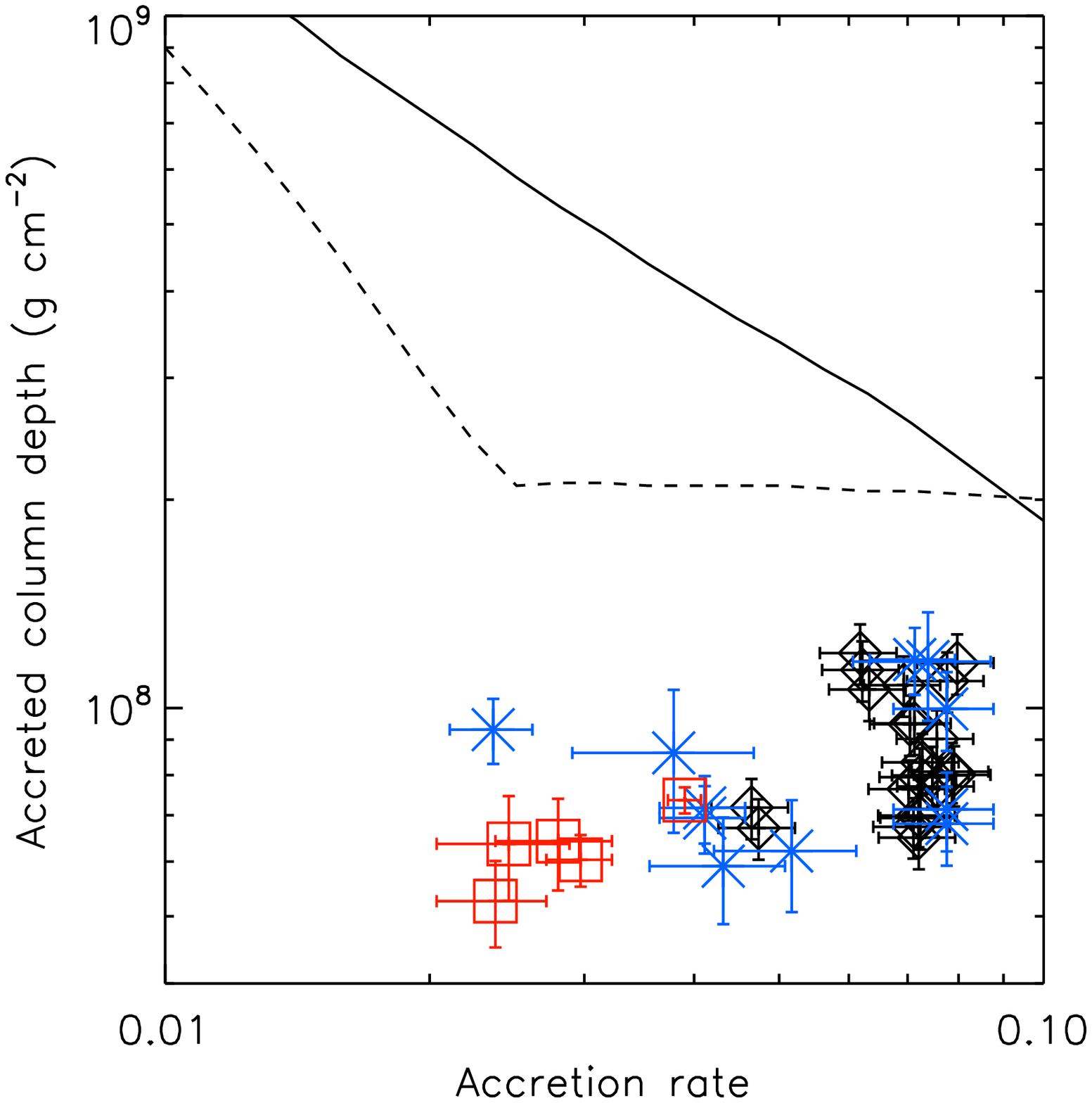}
\includegraphics[height=6.5cm,angle=0]{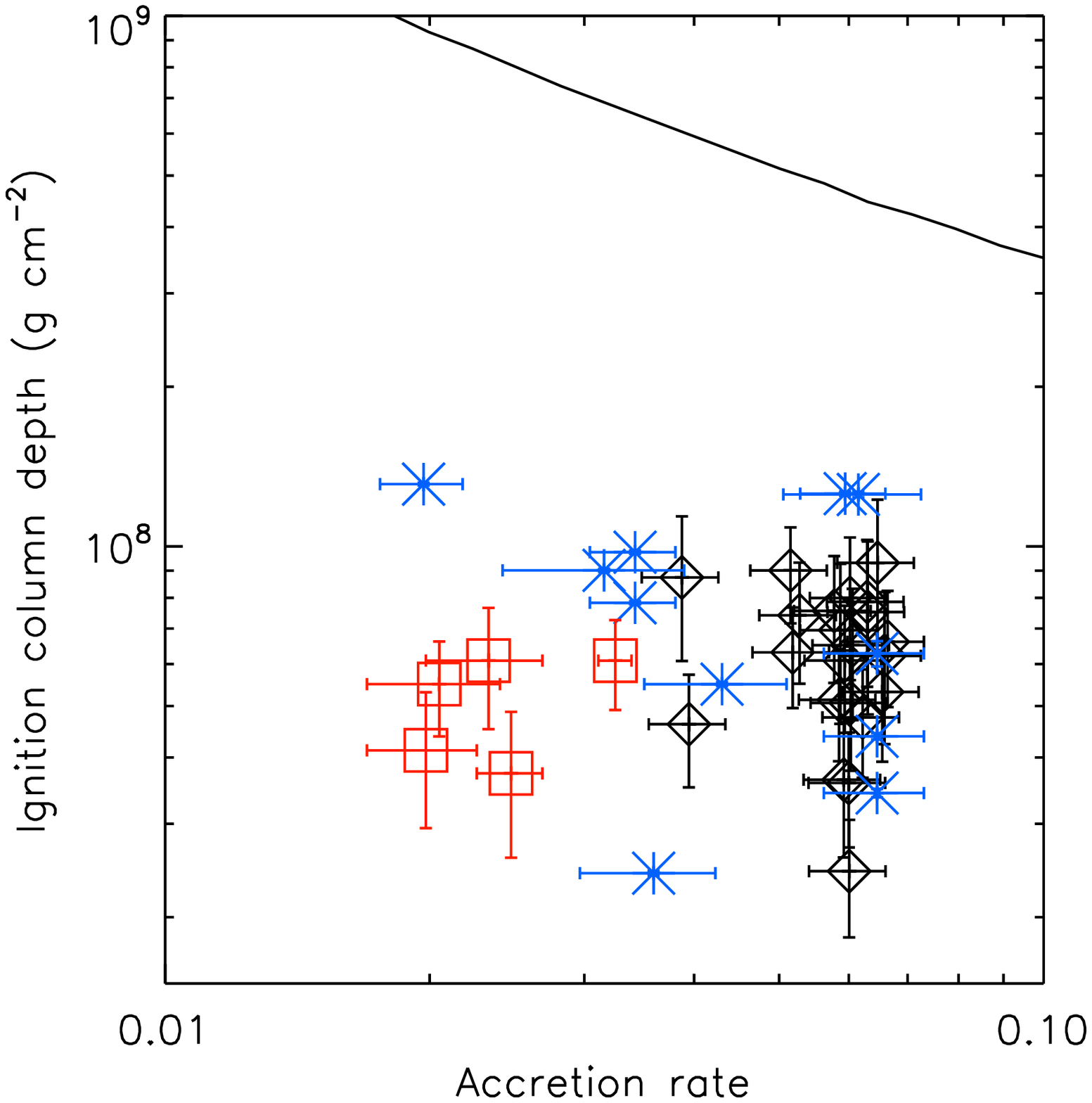}\includegraphics[height=6.5cm,angle=0]{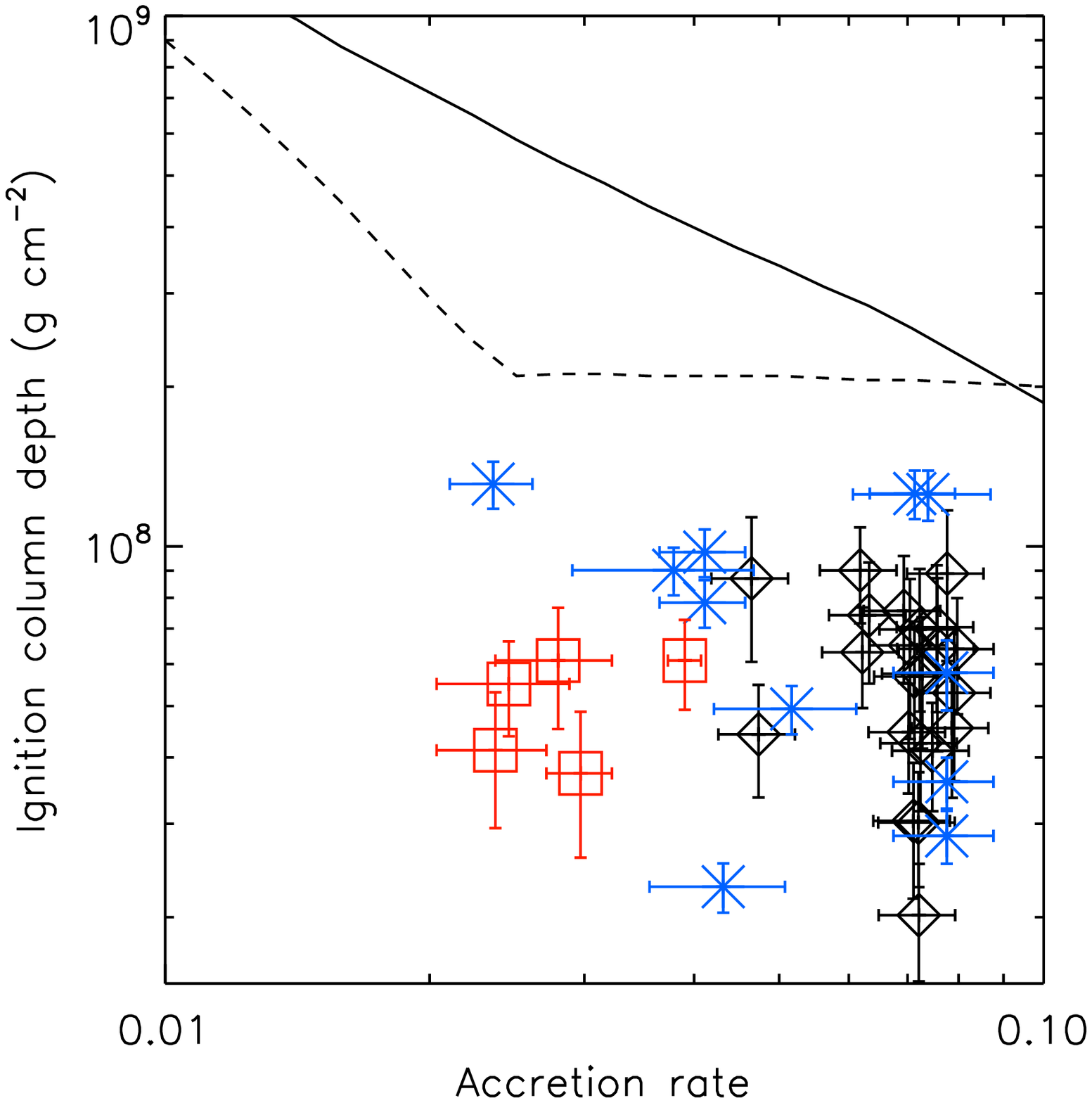}
\end{center}
\caption{{\bf Left:} The burst recurrence time (top), the corresponding accreted  column depth (calculated by multiplying the local accretion rate by the burst recurrence time; middle) and burst ignition column depth (calculated from the burst fluences; bottom) plotted as a function of the accretion rate. The
  accretion rate is in units of the Eddington accretion rate assuming pure
  helium fuel ($X_{\rm 0} = 0$). The line shows the model: a slow,
 suppressed neutrino 
emission process in the core and  $Q_{\rm imp}=100$, assuming
  the accreted material to be pure helium. The error bars represent the data:
  {\sl Chandra} (diamonds), {\sl RXTE} (asterisks) and {\sl INTEGRAL} (squares). The mass accretion rates
  are inferred from the persistent fluxes measured in the energy bands of the
  three instruments (see text for details). {\bf Right:} The same plot for the
  accreted material containing 20\% of hydrogen. The corresponding
  Eddington accretion rate is reduced. The two models shown are again for the 
  slow, suppressed cooling neutrino process with 
  $Q_{\rm imp}=100$, but assuming the accreted material has $X_{\rm 0}=0.2$ and a Solar metallicity (full line), and a metallicity ten times less than the Solar value (dashed line). 
\label{fig-Mdot-time-models}}
\end{figure}

\clearpage

Fig.~\ref{fig-Mdot-time-models} (left) shows the model and the observed 
 burst recurrence time and ignition column depths (calculated from the accreted mass and also from the mass consumed during the burst) plotted versus
 the mass accretion rate for
 {\sl Chandra}, {\sl RXTE} and {\sl INTEGRAL} data
assuming the accreted material to be pure helium. 
  Although the selected model maximizes the outward heat flux that ignites the
bursts, the comparison of the observed and theoretical burst recurrence times
 shows a large
discrepancy. For example, for a typical accretion rate of $\approx 0.05
\dot{M}_{\rm Edd}$ we observe bursts separated by 3$-$4 hours, while the model
predicts almost 15 hours.
To match the observed burst recurrence times  each of our estimated accretion rates would have to be multiplied by a correction factor in the range 2.5--6 (mean 3.6).  
 Such a large discrepancy may suggest that the region at the base of the
ignition column is significantly hotter than predicted by the model,
 which means
either that the cooling processes in the crust and core are slower than the
current theoretical lower limits, or that there is an additional heat source.

The additional heat may be supplied by burning of hydrogen, if the accreted
material is a mixture of hydrogen and helium. Since the stellar evolutionary
 models predict
that a small fraction of hydrogen   might be present in  evolved,
ultracompact binary systems, we produced a  model assuming the same core
 and crust
cooling properties, but with the accreted material that includes around
 20\%  of
hydrogen ($X_{\rm 0}=0.2$), which is a reasonable upper limit for an ultracompact binary.
Since the hydrogen burning is affected by the CNO fraction, we produced two
 models, one with the Solar CNO fraction of 0.012, and the other with the CNO
 fraction ten times lower (Z=0.0012). 
We also re-calculated the Eddington accretion rate limit for the new fuel
composition, and the new models are shown  in Fig.\ref{fig-Mdot-time-models}
(right).  Although the discrepancy between the model and data
 was slightly reduced, adding 20\% of hydrogen was not sufficient
to match the theoretical and the observed burst recurrence times. The model
with the low CNO fraction predicts  shorter  burst recurrence times. This is
expected, since the assumed hydrogen fraction of $X_{\rm 0}=0.2$ needs more than 30
hours to burn stably when the metallicity is 10 times lower than the Solar CNO
fraction (see Section~\ref{fluences}). Hence, the hydrogen will burn
stably for longer time and ignite later, at a larger depth and hence closer to
the helium burning region\footnote{This is not the case, however for very large mass accretion rates, because the ignition of helium occurs at much lower column depths}.

To estimate the ``missing heat'' we  produced several models, which further
increase the outward heat flux. These models also assume the
slow cooling and low conductivity of the neutron star core and crust as before, but include
 additional heating.
As our first attempt, we increased the total heat produced deep in the crust by electron capture
and pycnonucear reactions from the current theoretical estimate of 1.9 MeV \citep{HaenselZdunik2008} to 3.0 MeV per nucleon. 
Although  this  slightly reduced the burst recurrence times, it was not enough to match the observed values.

Recently,  \citet{2009ApJ...698.1020B} investigated the cooling rates
of several quasi-persistent transients during the
quiescence. They found that, in order to fit the cooling
light curves of these sources following an outburst, an inward-directed heat
flux from the shallow outer crust was required. For several sources in their
sample, \citet{2009ApJ...698.1020B} estimated the strength of this shallow
heat source to be in the range of 0.7$-$1.1 MeV per nucleon. Although the nature of this
shallow  heat
source is not yet clear, \citet{2009ApJ...698.1020B} suggest that its strength
 probably depends on the accretion rate.

Motivated by this result, we included an additional shallow heat source in our model.  For a strength of 0.7 MeV per nucleon, the additional heat source's effect was larger than that from the stronger heat produced deeper in the crust: it reduced the burst recurrence time from $\approx 10$ hours to $\approx 6.5$ hours (for $\dot{M} \approx 0.07 \dot{M}_{\rm Edd}$), approximately two times longer than observed.  However, regardless of the heat source's strength, the predicted ignition depth was always $> 1 \times 10^8$ g cm$^{-2}$, which is well above the observed values (Fig.~\ref{fig-Mdot-time-models}). Indeed, we find that helium burns stably for a sufficiently large assumed strength of the shallow heating.  The reason is the following \citep[see, e.g.,][]{1983ApJ...264..282P,1987ApJ...323L..55F}.
 Increasing the heat source's strength raises the burning layer's temperature, which decreases the column depth at which helium burning is thermally unstable \citep[e.g.,][]{1998mfns.conf..419B}. However, helium must survive to that ignition depth; increasing the temperature raises the burning rate and hence decreases the helium depletion depth.  If the former exceeds the latter, helium burns stably.



\section{Discussion}
\subsection{Comparison with the ignition model}

The most striking feature of the thermonuclear bursts exhibited by 4U~1728$-$34, apart from their consistently short rise and decay times, is the short recurrence times compared to the predictions of the ignition model. 
The simplest way to explain this discrepancy is that the measured accretion rates are underestimated, for example if additional soft or hard spectral components are present outside the instrumental bands
\footnote{Another possibility is that the assumed distance of 5.2 kpc, measured from the photospheric radius expansion bursts by \citet{2008ApJS..179..360G}, is underestimated. Although no other independent estimates of the distance are available for this source, we note that the assumed Eddington limit is close to the empirical critical luminosity measured by \citet{2003A&A...399..663K}. In addition, to match the model, the distance to 4U~1728$-$34 would have to be underestimated by a factor of 2--3.}.
 This is certainly plausible for the very narrow {\it Chandra}\/ band, but additional hard components are not possible for the {\it RXTE}\/ or {\it INTEGRAL}\/ measurements, unless they occur at very high energy ($\gtrsim100$~keV). Such features have not been detected in LMXBs. Low energy spectral components appear more likely, and could easily be overlooked thanks to the combination of low-energy sensitivity limits for {\it RXTE}\/ and {\it INTEGRAL}\/ of 2--3~keV, the restriction of the {\it Chandra}\/ spectral fits to $>1.5$~keV, and the strong absorption towards 4U~1728$-$34. However, we note that similar discrepancies were found for 4U~1820$-$30
 \citep{2003ApJ...595.1077C,2006ApJ...646..429C}, which is also well-studied and has low line-of-sight absorption ($N_H=$ a few times $10^{20}\ {\rm cm^{-2}}$; \citealt{2009ApJ...707L..77M}). Studies of the burst recurrence time -- persistent flux relationship in GS~1826$-$24 indicate that soft spectral components contributing up to 40\% of the broadband persistent flux can be present
\citep[e.g.,][]{2008ApJ...681..506T}. However, we note that for 4U\,1728$-$34 this additional flux would have to be typically 3.6 and up to 6 times larger than our measured bolometric fluxes (see Section~5), which seems implausible. Furthermore, unless the burst flux was also underestimated by a similar degree, such inflated luminosities would lead to intrinsic $\alpha$ values that were higher by the same factor, presenting additional difficulties. 

An obvious way to increase the burst ignition columns, without increasing the
global accretion rate, would be to restrict the neutron star area on which the
nuclear burning takes place.  The idea that  
strong magnetic fields could confine the accreted matter to a
fraction of the NS surface was first suggested by
 \citet{1980ApJ...238..287J}. \citet{1982ApJ...256..637A}
 later expanded on this idea \citep[see also][]{1997ApJ...477..897B,1998ApJ...496..915B}  although 
 the  observationally-inferred dipole magnetic fields of
 $\approx 10^8-10^{10}$ G \citep[e.g.,][]{1995xrbi.nasa..175L,1999A&AT...18..447P} seem to be much
lower than those required for the magnetic confinement.

Recently, \citet{2009ApJ...706..417L} proposed a model to explain the observed properties of the accretion-powered oscillations detected in the accreting millisecond X-ray pulsars (AMXPs). According to this model, the accreted material is channeled to the regions (spots) on the NS surface near the magnetic poles, which are close to the rotation axis (i.e., the magnetic and rotational axes are nearly aligned). \citet{2009ApJ...706..417L} argue that the mechanism that produces the nearly aligned axes  could significantly reduce the dipole component of the stellar magnetic field, without reducing the the field's total strength, which could be as high as $10^{11}-10^{12}$ G (i.e., strong enough to confine the accreted material).
 Furthermore, the  changes in  $\dot{M}$ and the  inner accretion disk
structure alter the size and location of 
 the X-ray emitting spots,
 which explains   the low (sometimes undetectable) pulse
amplitudes, nearly sinusoidal waveforms and  other properties of the
accretion-powered oscillations from AMXPs.

The effect of the accreted mass confinement on the burst ignition needs to be modelled in detail taking into account various effects \citep[e.g., the non-radial thermal diffusion is zero in the spherically-symmetric case but nonzero in the confinement case; see, e.g.,][]{2006ApJ...652..597P}. Here, we make a simple assumption that the ignition condition will not change significantly if the accretion is restricted to only part of the neutron star surface, and investigate if the simple area scaling  would be plausible for 4U~1728-34.

\begin{figure}
\begin{center}
\includegraphics[height=8cm,angle=0]{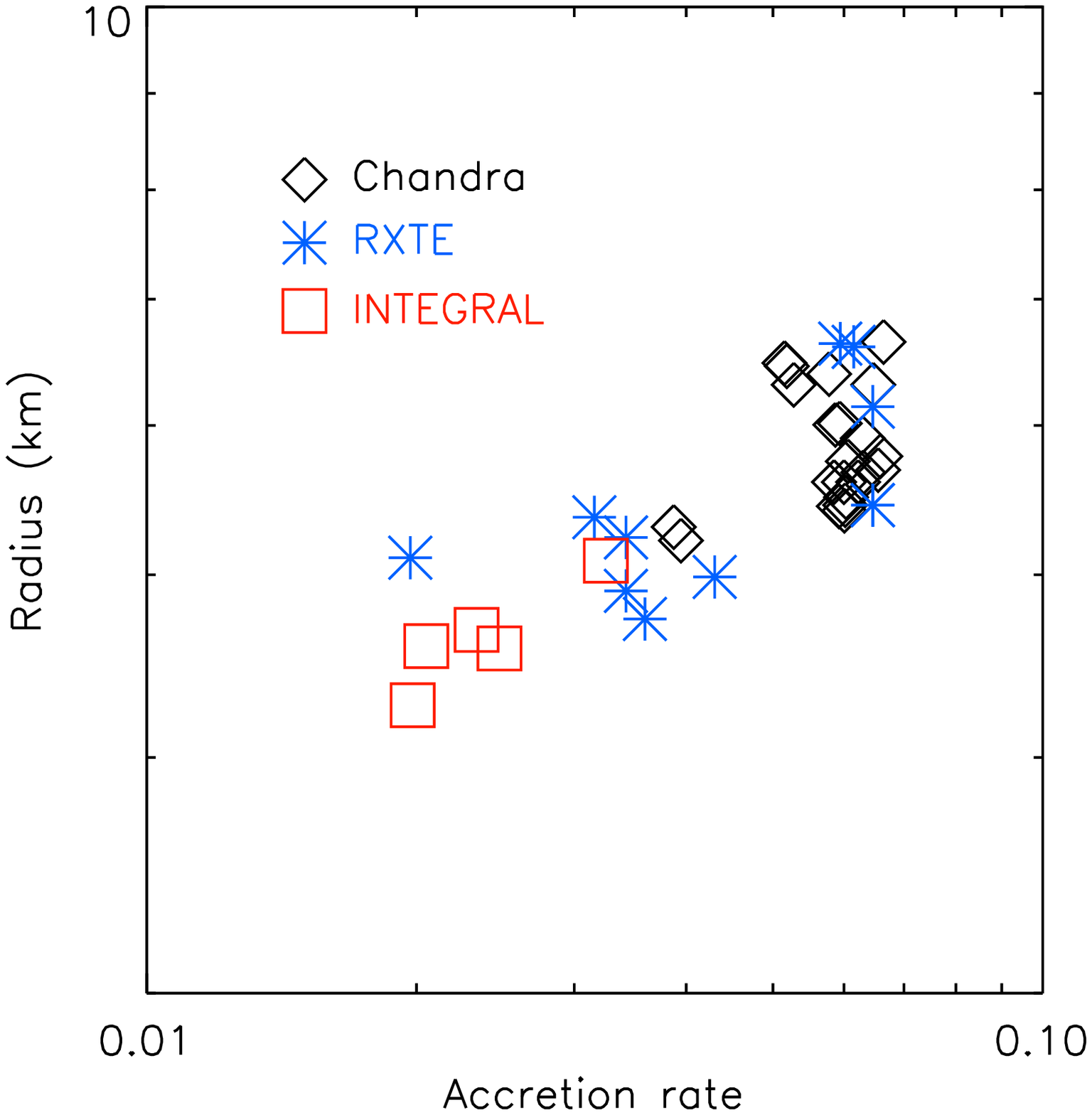}
\includegraphics[height=8cm,angle=0]{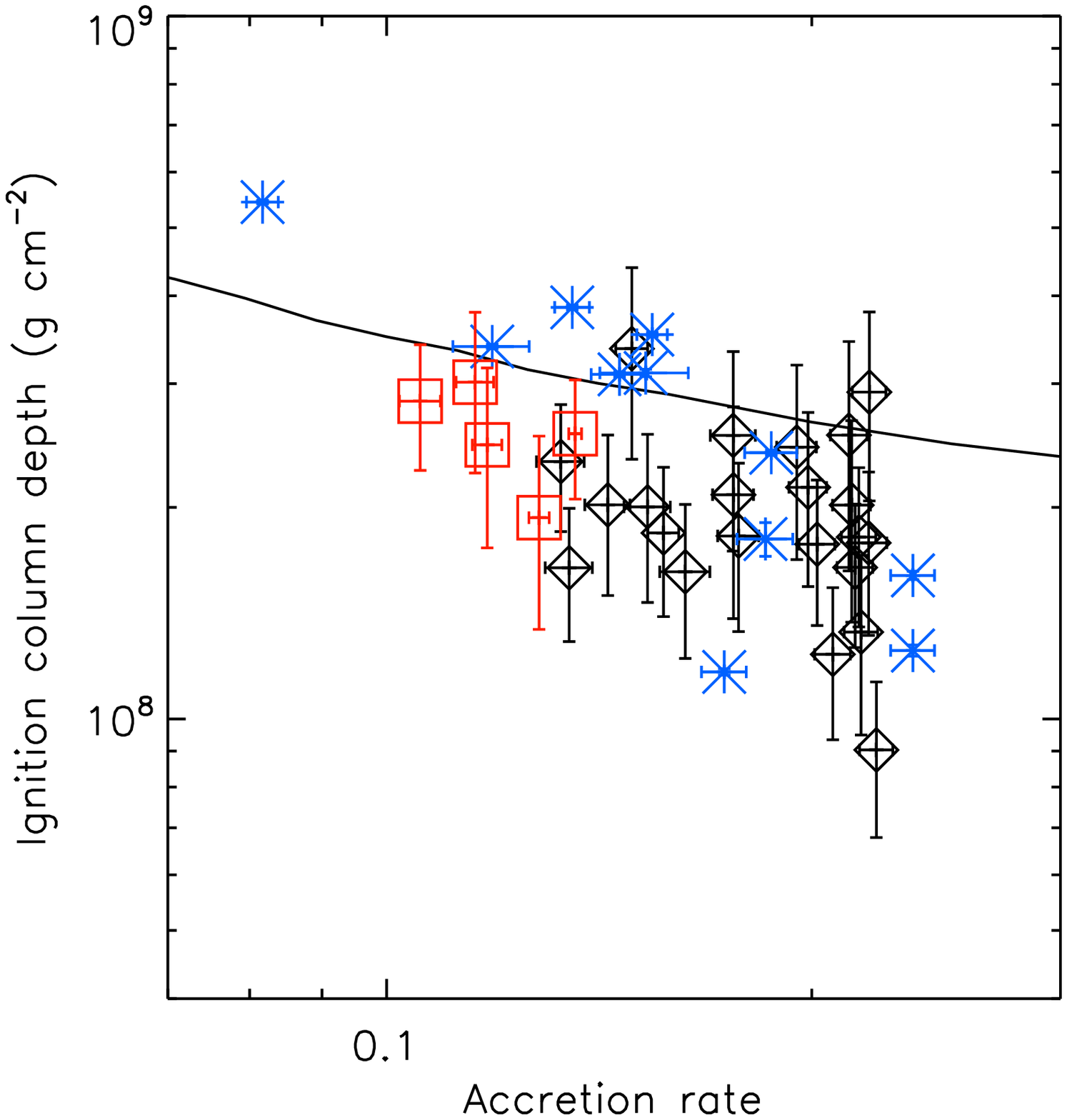}
\end{center}
\caption{{\bf Top:} The inferred radius of the neutron star area covered by the accreted material
plotted versus the observed global accretion rate. The assumed neutron star radius is 10.4 km (Section~4) while the covered area was calculated to match the predicted burst recurrence times.  {\bf Bottom:} The theoretical (full line) ignition column (calculated from the accretion rate; see also the middle panels of Fig.~\ref{fig-Mdot-time-models}) and the observed (1$\sigma$ error bars) ignition columns (measured from the burst fluences; see the bottom panels of Fig.~\ref{fig-Mdot-time-models}) plotted versus the scaled accretion rate. The scaling factors for each measurement are the same as in the left panel.   
\label{fig-area}}
\end{figure}

Fig.~\ref{fig-area} (left) shows the radius of the restricted area, which would be required so that the recurrence time (i.e., the accreted column) would match our model shown on the left panels in Fig.~\ref{fig-Mdot-time-models}.  The radius of the restricted area  changes from approximately 4.5$-$6 km, which seem to be  consistent with the measurements of the black-body emitting regions from the time-resolved burst spectra 
\citep[e.g., see Fig.~\ref{fig-time-resolved-spectra}; see also Fig.~5 in][]{2008ApJS..179..360G}. Fig.~\ref{fig-area} also suggests that  the size of the restricted area   may be correlated with the global accretion rate. 
This  is predicted by the moving spot model of  \citet{2009ApJ...706..417L}, suggesting that, perhaps, the nuclear burning in 4U~1728-34  is also concentrated to  moving regions  near the spin axis.

In general, a likely consequence of confinement of the nuclear burning is
that strong oscillations should be observed during every burst. However,
 burst oscillations were detected in only six out of 11 {\sl RXTE} bursts included in our sample \citep[for more detail see][]{2008ApJS..179..360G}.
The absence of burst oscillations during some bursts may also be explained
 if the burst oscillations also arise
from burning restricted to a region centered close to the spin axis and the latitude of this region varies slightly from burst to burst. 
Thus, the absence of strong burst oscillations during some of the bursts
observed by {\sl RXTE} from 4U~1728$-$34 does not necessarily rule out
fuel confinement.

In addition to confining the accreted fuel to a smaller area, the area scaling would also work if we assume a neutron star with a smaller radius. In addition to accumulating the fuel over a smaller area, we also expect the additional effects (e.g., the higher gravity makes the pressure higher for a given column depth, and the higher pressure usually gives a higher burning rate, etc.). However, based on our calculation for the neutron star radius of 6 km, these additional effects are not significant.

\subsection{The observed variations in the $\alpha$ parameter}

Here we address the problem of the large scatter of the $\alpha$ values  measured for 4U~1728-34 (see Figs.~\ref{fig-ratio-accRate} and \ref{fig-Mdot-time-models}).  
The right panel of Fig.~\ref{fig-area} shows  the ignition column scatter after the area scaling. The model column and the column calculated from the measured accretion rate coincide, after scaling   by the appropriate area
  (shown on the left panel). However,  the ignition column calculated from the observed burst fluences and scaled by the same area  shows a scatter.

 Similar large $\alpha$ deviations are observed in other X-ray bursters \citep[e.g.,][and references therein]{1988ApJ...324..995F}, and a model to explain it was suggested by \citet{1987ApJ...319..902F}. According to their model, the observed large $\alpha$ deviations from burst to burst can be explained by a fuel buffer in the envelope.  
 If the size of the fuel buffer is comparable to the total energy of the burst
 (which could easily be the case for the weak bursts observed from 4U~1728-34)
 the $\alpha$ deviations could be large. However, as suggested by
 \citet{1987ApJ...319..902F}, the influence of the buffer can be reduced
 statistically if a large number of bursts is observed.  
If this mechanism is responsible for the observed $\alpha$ scatter, for
  consecutive bursts we
  would expect a pattern of alternating high and low $\alpha$ values following
  each other.
 Although our sample includes 13 consecutive {\sl Chandra}
  bursts, the large $\alpha$ uncertainties may have prevented us to observe this behavior. The
  $\alpha$ uncertainties of the {\sl RXTE} bursts are significantly smaller,
but our sample does not include more than three consecutive {\sl RXTE} bursts.

The mean value of the $\alpha$ parameter measured from the 38 bursts in our sample is 180, and we do not see any variations of the mean value with the accretion rate, at least in our observed range of the accretion rates from $0.01-0.07~\dot{M}_{\rm Edd}$. This is close to the value predicted by our model (180-190; see Fig.~\ref{fig-ratio-accRate}), which seems to include a small ``leak'' (less than 10\%) due to some stable burning before each burst.  Such a close agreement between the average  and theoretical $\alpha$ values implies that the accretion disk inclination angle with respect to the line of sight is approximately 60 degrees (or slightly less for $X= \langle 0.2 \rangle$), as predicted 
in the model of angular distribution of radiation
 by \citet[][see their Fig.~2 and formula 6]{1988ApJ...324..995F}. The model suggests that the emission is enhanced in the direction perpendicular to the accretion disk plane, due to scattering in the inner disk, and the degree of anisotropy is larger for the persistent emission. 
The angle of $\approx$60 degrees is consistent with the fact that the eclipses
were never observed from this source. \citet{1987ApJ...319..902F} also argue
that the $\alpha$ deviations are largest for the photospheric radius expansion
bursts (PRE) because part of the outer envelope can be ejected,
increasing the $\alpha$ significantly. Our sample includes four possible 
  PRE bursts detected by {\sl Chandra} (see Galloway et al.~2010) and also two
 PRE bursts in the {\sl RXTE} sample \citep{2008ApJS..179..360G}, but
their $\alpha$ parameters are not exceptional. This may suggest that the
 mass loss due to
the envelope ejection during PRE bursts is negligible compared to the
fuel buffer size \citep[e.g., see][]{2006ApJ...639.1018W}.

The same mechanism that pushes the envelope buffer closer to the ignition
zone  \citep[i.e., the turbulent mixing and  dissipative heating associated
with thermonuclear instabilities, caused by inflow of angular momentum with
accreted gas; ][]{1987ApJ...319..902F} could also explain weak, frequent
bursts (``premature ignition''). A similar idea was also proposed recently by
\citet{2007ApJ...663.1252P}. The turbulent mixing model alone could, perhaps,
explain the observed burst properties, without introducing the accretion
confinement, but although its spin is slow \citep[363 Hz;][]{1996IAUC.6387....2S}, we suspect that the observed
accretion rates of 4U~1728$-$34 might be too small for this mechanism to
work. According to this model, the mixing is most effective for large accretion rates and small
  spin rates.  Figure 15 of \citet{2007ApJ...663.1252P} suggests that for
  the slow spin of 4U~1728$-$34, the
required accretion rate to overcome the buoyancy barrier would be
 $\gtrsim 0.1 \dot{M}_{\rm Edd}$. 
However, the critical accretion rate depends sensitively on the conditions in the fuel layer and the strength of mixing \citep[e.g., see equation 59 of][]{2007ApJ...663.1252P}.

It is worth noting that our sample is highly selective for bursts with
short recurrence times; the bursts observed with {\it Chandra}\/ all
have unusually short recurrence times, and for {\it RXTE}\/ and {\it
INTEGRAL}\/ we only quote measurements for pairs of bursts for which
there are no (or few) data gaps. If we relax this requirement, there is
evidence for much higher (up to $\approx1000$) values of $\alpha$ in the
literature for 4U~1728$-$34 and other systems. \citet{2006A&A...458...21F}
quote $\alpha$ values up to almost 800, but these are upper limits
only, since one or more data gaps occurred in between the bursts.
However, \citet{2008ApJS..179..360G} also found that the average $\alpha$-value for a group of
bursts with short recurrence times (including 4U~1728$-$34) was
$>1000$. Such high $\alpha$ values likely cannot arise from incomplete
burning leaving behind a residual fuel buffer, since the required size
of the buffer would be many times the amount of fuel burnt in the
burst. Instead, such inefficient burning must arise from a more
substantial energy leak, perhaps steady He-burning, as has also been
suggested to play a role in the decrease in burst rate for most
bursters at an accretion rate of around $0.1\,\dot{M}_{\rm Edd}$ 
\citep{2001A&A...372..138R,2003A&A...405.1033C,2008ApJS..179..360G}.
Some degree
of steady burning could also explain the scatter in the measured 
$\alpha$ values in our sample, but we cannot rule out the other
mechanisms discussed here.

\subsection{Comparison with other bursters}

Finally, we investigate whether the accreted fuel confinement proposed to
  explain the observed burst recurrence times of 4U~1728--34, and other
  mechanisms discussed here, may also be applicable to other
  bursters.
In their study of 9
bursters observed with {\sl BeppoSAX}, \citet{2003A&A...405.1033C} found that
the burst rate increases slowly up to the peak rate, reaching it at
approximately $1-2 \times 10^{37}$ erg s$^{-1}$ (or $\sim0.05-0.1 \dot{M}_{\rm Edd}$  for mixed fuel), after which it sharply
drops. \citet{2003A&A...405.1033C} attributed such a behavior to the onset of
stable hydrogen burning, which occurs at the transition from the lowest to the
medium accretion rates \citep[i.e., from the burst ignition regime 3 to 2;
][]{1981ApJ...247..267F}.

 The drop in the observed burst rate, at a slightly
lower X-ray luminosity, was also later confirmed by \citet{2008ApJS..179..360G}, who
studied a much larger sample of bursters covering a much wider range of
accretion rates. However,
\citet{2008ApJS..179..360G} argue that the decreasing burst frequency is more likely  due to
onset of stable burning of helium \citep[see also][]{1988MNRAS.233..437V}, which must  occur at lower accretion rates
than theoretically predicted. \citet{2008ApJS..179..360G} also found that the burst rate 
continues to decrease further until the bursting entirely stops
 (at $\sim0.3 \dot{M}_{\rm Edd}$). In addition, the $\alpha$ parameter was found to remain constant or to
slightly increase globally, rather than decrease as would be expected at the
transition from the smallest accretion rate (ignition regime 3) to the medium accretion
rate regime (2) and onset
of stable burning of hydrogen \citep[i.e., the transition of the H/He ignition
to the He-ignition;][]{1981ApJ...247..267F}.

Interpreting their results, \citet{2003A&A...405.1033C} pointed out that there
 is a large discrepancy between the theoretically predicted ($\sim0.01 \dot{M}_{\rm Edd}$) and measured
boundary between the two regimes, and offered an alternative
explanation, which was proposed earlier by \citet{2000AIPC..522..359B} based on the
\citet{1999AstL...25..269I} model. \citet{2000AIPC..522..359B} suggested
 that the observed burst rate drop is caused by the decrease of the local
 accretion rate. This would be possible if the nuclear burning is restricted
 to a smaller area, which increases  with the global accretion
 rate. If the area increases faster than the global accretion rate, the local
 accretion rate may be decreasing. The correlation  between  the black-body  emitting
 radius, measured in the burst cooling tail of EXO~0748--676 \citep{1986ApJ...308..213G}, and the
 corresponding accretion rate supports this hypothesis. The observed
  $\alpha$ increase, following the burst drop
  \citep{2008ApJS..179..360G}, is also consistent with the \citet{2000AIPC..522..359B}
  interpretation (in this case the transition would be from
  the highest accretion rate regime (1) to regime 2, rather than from 3 to 2).
Although the fractional covering of the accreted fuel
may not be required to explain the bursting behavior of all sources, it
certainly seems plausible at least for some
bursters.

Finally, we note that the same mechanism, which we believe is responsible
 for the
large $\alpha$ scatter measured for 4U~1728-34 \citep[i.e., the fuel storage and its
ignition due to mixing;][]{1987ApJ...319..902F}, is also the most likely
explanation for the large $\alpha$ scatter and weak frequent bursts with very
short 
recurrence times of 10 minutes, observed in bursters accreting mixed fuel
\citep[e.g., 4U~1636--53, EXO~0748--676;][]{1987ApJ...319..893L,1986ApJ...308..213G,1987ApJ...323..575G}.
However, the scatter in $\alpha$ for the latter sources (10--500 for 4U~1636--53, and 5--50 for  EXO~0748--676) is somewhat larger than that measured for 1728-34, likely because the fuel composition for sources accreting mixed H/He can vary due to steady H-burning prior to ignition. Interestingly, while our results suggest that some fraction of the burst fuel is typically left unburnt in 4U 1728-34, short recurrence time bursts are not observed in this source, whereas the same mechanism is thought to give rise to such bursts in systems accreting mixed H/He \citep{keek}.

\section{Conclusions}
\label{conclusions}

We compared the observed properties of 38 bursts detected in {\sl Chandra},
{\sl RXTE} and  {\sl INTEGRAL} observations of the helium-rich accretor
4U~1728$-$34 with new ignition models.
We find that the observationaly-inferred
ignition depths, assuming complete fuel spreading on the neutron star surface,
are significantly smaller than the theoretically-derived minimum possible
ignition depth of $1-2\times 10^8$ g cm$^2$.

One way to reconcile the observed and predicted burst recurrence times would be to assume that the observed X-ray luminosities underestimate the accretion rates (for example, due to a non-detection of the  extremely soft or/and hard spectral components). However, for 4U\,1728$-$34 this scenario is not plausable because these additional spectral components would have to be 2$-$6 times larger than the mesured bolometric luminosity.  In addition, it would imply significantly larger $\alpha$ values than theoretically predicted.

Alternatively, the ignition column could be increased, without increasing the global accretion rate,  if we assume that the accreted material is confined to a resticted area  on the neutron star surface. 
To match the observations with our ignition model, we find that the spot radii would have to be in the range  
 4.5$-$6 km, and that they seem to be correlated 
 with the global accretion rate as predicted by the \citet{2009ApJ...706..417L}  model. However, detailed ignition models that include all possible effects of the magnetic confinement are needed to confirm this result. An additional confirmation could come from the comparison of the X-ray emission from the accretion disk with the nuclear oscillation  amplitudes detected simultaneously from the source.  

To explain the weak, frequent bursts observed from 4U~1728$-$34, we also consider shear-triggered mixing of the accreted helium to larger column depths \citep{1987ApJ...319..902F,2007ApJ...663.1252P}. However, while this mechanism could push the fuel buffer from the outer envelope closer to the burning zone and explain the observed large $\alpha$ deviations, it might not be sufficient to explain the observed burst properties alone
  at the relatively low observed accretion rates.

{\it We would like to thank Ed Brown, Maurizio Falanga and Stratos Boutloukos
  for helpful discussions. We would also like to thank the anonymous referee
  for the useful and constructive comments. }

\bibliographystyle{apj}
\bibliography{./paper}

\end{document}